\documentclass[floatfix,aps,prb,pdflatex,reprint,10pt,showpacs,superscriptaddress]{revtex4-2}
\usepackage{float}

\usepackage{amssymb}
\usepackage{amsthm}
\usepackage{bbm}
\usepackage{ulem}
\usepackage{graphicx}
\usepackage{hyperref}
\usepackage{physics}
\usepackage{systeme}
\usepackage{times}
\usepackage{epstopdf}
\usepackage{float}
\usepackage{xcolor}
\usepackage{booktabs}
\usepackage{cancel}
\usepackage{mathtools}

\hypersetup{
colorlinks=true,linkcolor=blue,citecolor=blue,
filecolor=blue,urlcolor=blue,breaklinks=true
}

\RequirePackage{color}

\begin{document}

\date{\today}

\title{Quantum Hall-like effect for neutral particles with magnetic dipole moments in a quantum dot}

\author{Carlos Magno O. Pereira}
\email[Carlos Magno O. Pereira - ]{carlos.mop@discente.ufma.br}
\affiliation{Departamento de F\'{\i}sica, Universidade Federal do Maranh\~{a}o, 65085-580 S\~{a}o Lu\'{\i}s, Maranh\~{a}o, Brazil}

\author{Edilberto O. Silva}
\email[Edilberto O. Silva - ]{edilberto.silva@ufma.br}
\affiliation{Departamento de F\'{\i}sica, Universidade Federal do Maranh\~{a}o, 65085-580 S\~{a}o Lu\'{\i}s, Maranh\~{a}o, Brazil}

\begin{abstract}
We predict a new class of quantum Hall phenomena in completely neutral systems, demonstrating that the interplay between radial electric fields and dipole moments induces exact $e^2/h$ quantization without Landau levels or external magnetic fields. Contrary to conventional wisdom, our theory reveals that: (i) the singularity of line charges does not destroy topological protection, (ii) spin-control of quantization emerges from boundary conditions alone, and (iii) the effect persists up to 25 K, surpassing typical neutral systems. These findings establish electric field engineering as a viable route to topological matter beyond magnetic paradigms.
\end{abstract}

\maketitle

\section{Introduction}

The quantum Hall effect (QHE), first observed in two-dimensional electron systems under strong magnetic fields, represents one of the most profound manifestations of topological order in condensed matter physics. The discoveries of both integer~\cite{PRL.1980.45.494} and fractional~\cite{PRL.1982.48.1559} quantum Hall effects not only revolutionized our understanding of quantum phases but also set new standards for precision metrology through the resistance quantum $h/e^2$ \cite{NE.2024.7.443}. These breakthroughs naturally inspired the search for analogous phenomena in systems beyond charged particles, particularly in neutral quantum matter, where unconventional control parameters might give rise to similar topological protection.

The Aharonov-Casher (AC) effect~\cite{PRL.1984.53.319} provides a fundamental mechanism for such neutral analogs, where magnetic dipole moments coupled to electric fields can emulate key aspects of quantum Hall physics. This relativistic spin-electric coupling has been explored in diverse platforms ranging from cold atom systems~\cite{RMP.2011.83.1523} and neutron interferometry~\cite{PRA.2014.89.042101} to photonic crystals~\cite{Nature.2019.565.622}, polar molecular gases~\cite{PRL.2014.113.025301}, as well as in recent advances such as the magnonic Aharonov-Casher effect~\cite{PRB.2023.108.L220404,PRB.2025.111.174440}, spin wave optics~\cite{PRB.2024.109.184437}, Aharonov-Casher bound states \cite{EPJC.2013.73.2402,PRA.2024.110.062213}, and spintronic networks~\cite{PRB.2019.100.161108}.
These works highlight a growing interest in leveraging the AC effect across a wide array of physical media, albeit often requiring complex experimental conditions. However, most theoretical models require either idealized field configurations that are challenging to realize experimentally or complex setups involving crossed electric and magnetic fields~\cite{PLA.2017.381.849}, which limits their practical implementation. While the Aharonov-Casher effect in neutral particles has been investigated across various platforms, the closest work to our proposal is that in Ref.~\cite{PLA.2017.381.849}, which explored an analog of the Quantum Hall Effect for neutral particles with a magnetic dipole moment. However, their model relies on the presence of external, crossed, and inhomogeneous electric and magnetic fields, where the dipole's interaction with the magnetic field is an essential component for the Landau-like quantization. In contrast, our work proposes a distinct regime: the emergence of a quantized Hall-like effect in a quantum dot solely from the interaction of magnetic dipoles with a purely radial electric field, without the need for an external magnetic field. This simplification of the experimental configuration, utilizing only an electrostatic field, represents a novel and promising route for engineering topological matter in neutral systems. Addressing these challenges, in this work, we demonstrate that a straightforward configuration, a quantum dot subjected to the radial electric field of a charged wire, can support robust quantum Hall-like effects in neutral particles. This system reveals quantized transport signatures reminiscent of the Landau problem, without involving magnetic fields or nontrivial gauge constructions. The system's elegance lies in its minimal ingredients: the electric field $\mathbf{E} = (2\lambda/r)\hat{\mathbf{r}}$ from a line charge combines with the Aharonov-Casher interaction $\mathbf{A}_{\text{eff}} = \boldsymbol{\mu} \times \mathbf{E}$ to create Landau-level analogs, while harmonic confinement $V(r) = \frac{1}{2}M\omega_0^2r^2$ quantizes the spectrum. Crucially, this approach circumvents the need for magnetic fields or complex field geometries, relying instead on a single electrostatic element whose parameters are experimentally accessible in existing platforms.

We emphasize the experimental viability of our proposal by showing its compatibility with two well-established platforms. For semiconductor implementations, the required confinement energy $\hbar\omega_0 = 0.25$ meV and linear charge density $\lambda \sim 10^{-11}$ C/m fall squarely within the capabilities of electrostatically defined GaAs quantum dots~\cite{PRB.2000.62.7045}. Simultaneously, cold atom platforms can achieve equivalent conditions through synthetic gauge fields~\cite{NJP.2003.5.56,PRA.2009.79.063613}, where $\lambda$ becomes tunable via laser intensity. This dual realizability across different physical systems significantly enhances the prospects for experimental verification.

What distinguishes our work from previous studies is its treatment of three key challenges in neutral quantum Hall systems. First, the apparent singularity of the line charge field, embodied in the term $\boldsymbol{\nabla}\cdot\mathbf{E} = 2\lambda\delta(r)/r$, is resolved through careful application of self-adjoint extension methods~\cite{EPJC.2013.73.2402}, demonstrating that topological protection can survive even in the presence of such singularities. Second, the system achieves complete control over the sign of the Hall conductivity solely through spin orientation, without requiring external magnetic fields, a feature evident in the spectrum's dependence on the AC parameter $\xi = 2s\mu\lambda/\hbar$. Third, and perhaps most surprisingly, the quantization persists up to 25 K, far exceeding the temperature limits of typical neutral systems~\cite{Nature.2019.565.622} and approaching the operational range of conventional semiconductor quantum Hall devices.

The manuscript is organized to guide the reader through both theoretical foundations and experimental implications. Section~\ref{sec:model} establishes the effective Hamiltonian for neutral particles with magnetic dipole moments, and analyzes the emergent quantum spectrum, including its Landau-level analogies. Section~\ref{persistent_current} then derives the $e^2/h$-quantized Hall response through persistent currents. Throughout our discussion, we maintain a strong connection to experimental realizations in both semiconductor nanostructures~\cite{PRB.2000.62.7045} and cold-atom platforms~\cite{PRA.2009.79.063613}, providing a comprehensive roadmap for observing these effects in the laboratory. Our conclusions and additional comments are presented in Section \ref{conclusion}.

\section{Planar Dynamics and Effective Hamiltonian \label{sec:model}}

\begin{figure}[tbh]
\centering
\includegraphics[width=1.0\linewidth]{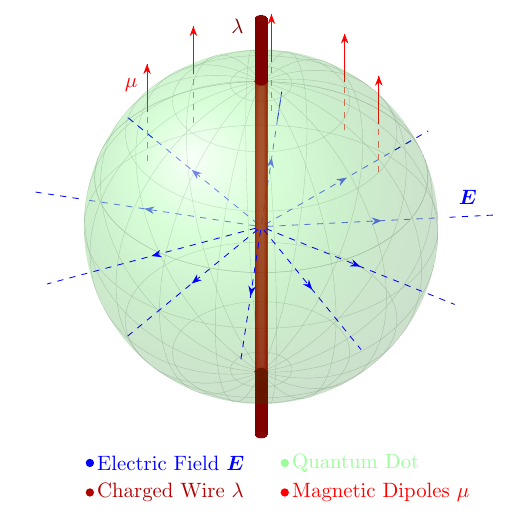}
\caption{Schematic representation of the proposed system: (a) A quantum dot (green region) with harmonic confinement potential $V(r) = \frac{1}{2}M\omega_0^2r^2$ surrounds a charged wire (red cylinder) generating the radial electric field $\mathbf{E} = (2\lambda/r)\hat{\mathbf{r}}$ (blue arrows). Magnetic dipoles $\boldsymbol{\mu}$ (red arrows) experience the effective gauge potential $\mathbf{A}_{\text{eff}} = \boldsymbol{\mu} \times \mathbf{E}$.}
\label{fig:system}
\end{figure}

To describe the planar dynamics of a neutral particle with a magnetic dipole moment $ \boldsymbol{\mu} $, we follow the formalism developed in Ref.~\cite{EPJC.2014.74.3112}. The magnetic moment is intrinsically linked to the spin of the particle. For a spin-1/2 particle, it is given by $\boldsymbol{\mu} = (g e/2 M)\mathbf{S}$,
where $g$ is the Landé g-factor, and $\mathbf{S} = (\hbar/2) \boldsymbol{\sigma}$ is the spin operator. Assuming a spin-polarized configuration with spin aligned along the $ z $-axis, we take $ \boldsymbol{\mu} = \mu \hat{\boldsymbol{z}} $, with $ \mu = \pm (g e \hbar/4 M)$. Thus, starting from the (3+1)-dimensional Dirac theory, one can reduce the Hamiltonian to its planar form by assuming motion confined to the $xy$-plane and static fields depending only on $r$. This allows a consistent separation of variables and cancellation of terms involving the third spatial direction. In the planar limit, the nonrelativistic Pauli Hamiltonian becomes \cite{PRL.1990.64.2347,IJMPA.1991.6.3119,EPJC.2014.74.3112,EPJC.2013.73.2402}
\begin{align}
H = \frac{1}{2M} \left(\mathbf{p} - \boldsymbol{\mu} \times \mathbf{E} \right)^2 + \frac{\mu}{2M}\boldsymbol{\nabla} \cdot \mathbf{E} + \frac{1}{2}M \omega_0^2 r^2, \label{hm}
\end{align}
where we have explicitly removed the longitudinal ($z$-direction) contributions, which cancel due to symmetry, and retained the singular contribution from the divergence of the electric field. The effective gauge potential in Eq.~(\ref{hm}) is given by $\mathbf{A}_{\text{eff}} = \boldsymbol{\mu} \times \mathbf{E} = (\xi \hbar/r) \boldsymbol{\hat{\varphi}}$, with $\xi$ encapsulating the spin contribution. Although our fundamental description considers generic neutral particles, this model can be adapted to semiconductor quantum dots (e.g., GaAs) by considering neutral excitations such as excitons or trions, where the net charge is zero, effective magnetic moments arising from spin-orbit interactions, and dressed electronic states that behave as neutral quasiparticles. For charged electrons in GaAs, additional terms would be required in the Hamiltonian to account for Coulomb interactions.

Figure~\ref{fig:system} illustrates our model: a two-dimensional quantum dot (e.g., a GaAs heterostructure) with a charged wire positioned centrally along the $z$-axis. The charged wire generates an electric field that induces a gauge potential that depends on the spin of the particle through the AC effect. The magnetic dipole moments of neutral particles confined by the harmonic potential acquire a geometric phase that plays the role of a vector potential of the magnetic field in charged systems. This configuration enables the emergence of Landau-type quantization without the need for magnetic fields, laying the groundwork for the study of quantum Hall physics in neutral matter.

At this point, we emphasize that the appearance of the term $\boldsymbol{\nabla} \cdot \mathbf{E} = 2\lambda \delta(r)/r$ requires a careful mathematical treatment. In this work, we follow the approach in Ref.~\cite{EPJC.2013.73.2402}, where the singular behavior at the origin is addressed using the self-adjoint extension method, allowing for consistent boundary conditions at $r = 0$. The resulting energy spectrum, accounting for the singularity, is found to be
\begin{equation}
E_{n,m} = \hbar \omega_0 \left( 2n + 1 \pm |m -\xi| \right).
\end{equation}
For our purposes, only the energy spectrum is required.
\begin{figure}
    \centering
    \includegraphics[width=1\linewidth]{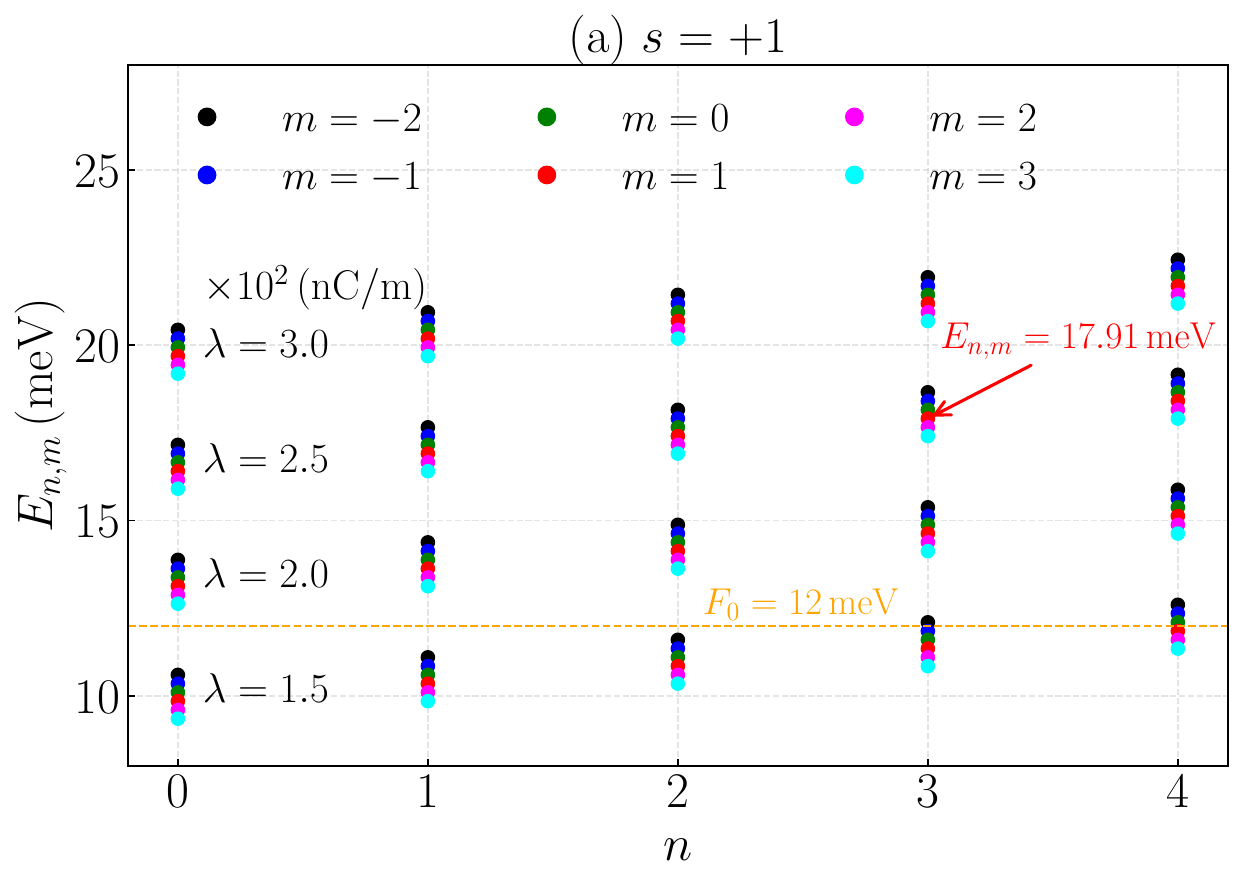}\\
    \includegraphics[width=1\linewidth]{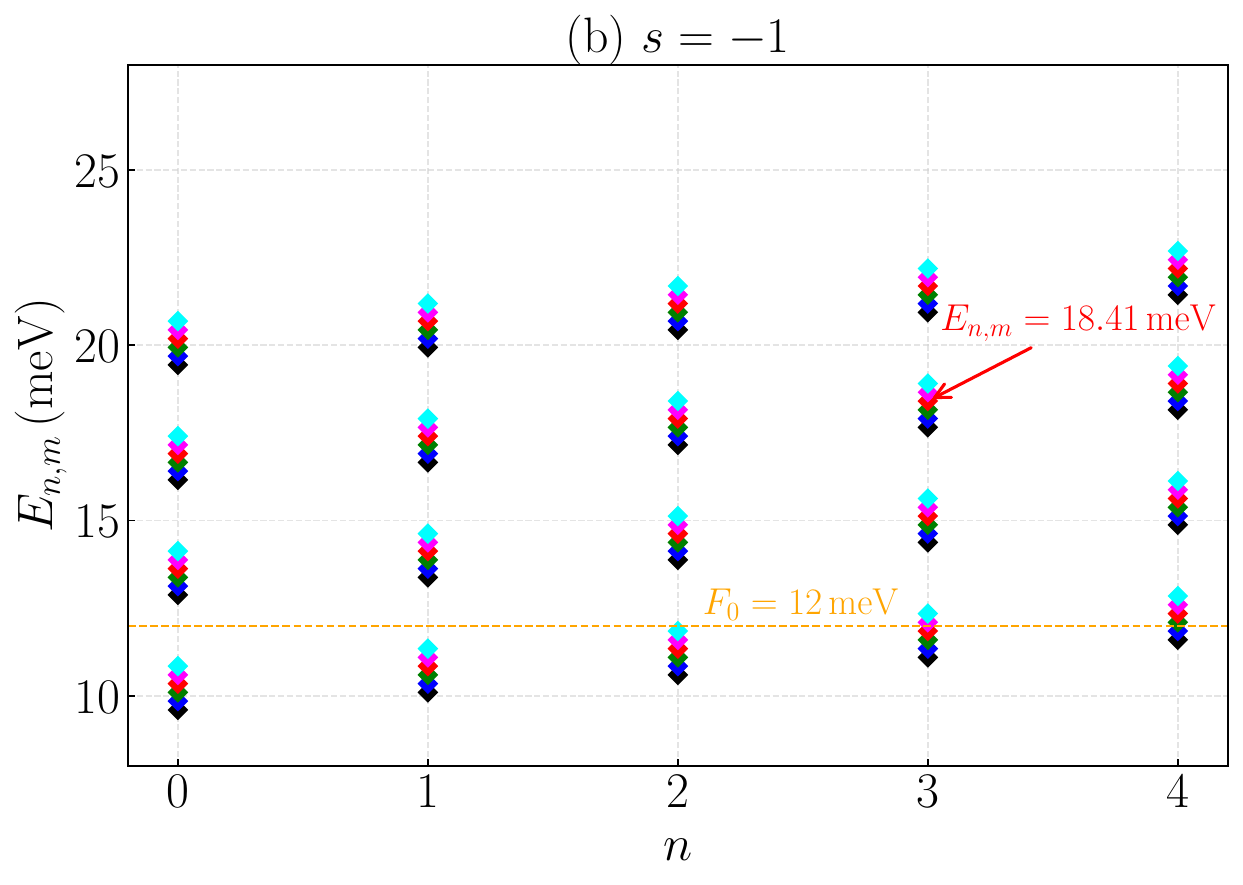}
    \caption{Energy spectrum $ E_{n,m} $ for cases (a) $ s = +1 $ and (b) $ s = -1 $, considering different principal quantum numbers $ n = 0, 1, \ldots, 4 $ and angular momenta $ m = -2, -1, \ldots, 3 $, with charge density values $ \lambda = 1{,}5,\, 2{,}0,\, 2{,}5,\, 3{,}0 \times 10^{2}\,\mathrm{nC}/\mathrm{m} $. Each color represents a distinct value of $ m $. The circle-shaped markers indicate the case $ s = +1 $ (spin up) and the diamond-shaped markers represent $ s = -1 $ (spin down).}
    \label{fig:energy-n}
\end{figure}

The Fig.~\ref{fig:energy-n} shows the energy spectrum $ E_{n,m} $ as a function of the principal quantum numbers $ n $ and angular momentum $ m $ for two distinct cases: (a) $ s = +1 $ (spin up) and (b) $ s = -1 $ (spin down). It can be seen that, for fixed values of $ m $ and $ \lambda $, the energy increases with increasing $ n $, as expected. Comparing the two panels, it can be seen that, for the same set of $ n $, $ m $ and $ \lambda $, the states with $ s = -1 $ have slightly higher energy levels than those with $ s = +1 $, evidencing the splitting due to spin-orbit coupling. In addition, the increase in charge density $ \lambda $ causes a progressive shift of the energy levels to higher values, moving away from the initial Fermi energy (12\,meV). It should also be noted that for $ n \leq 2 $ and $ \lambda = 1.5 \times 10^2 \,\mathrm{nC/m} $, the energy levels remain below the Fermi energy, indicating that these states are occupied in the low-density regime.

\begin{figure}[tbh]
\centering
\includegraphics[width=1.0\linewidth]{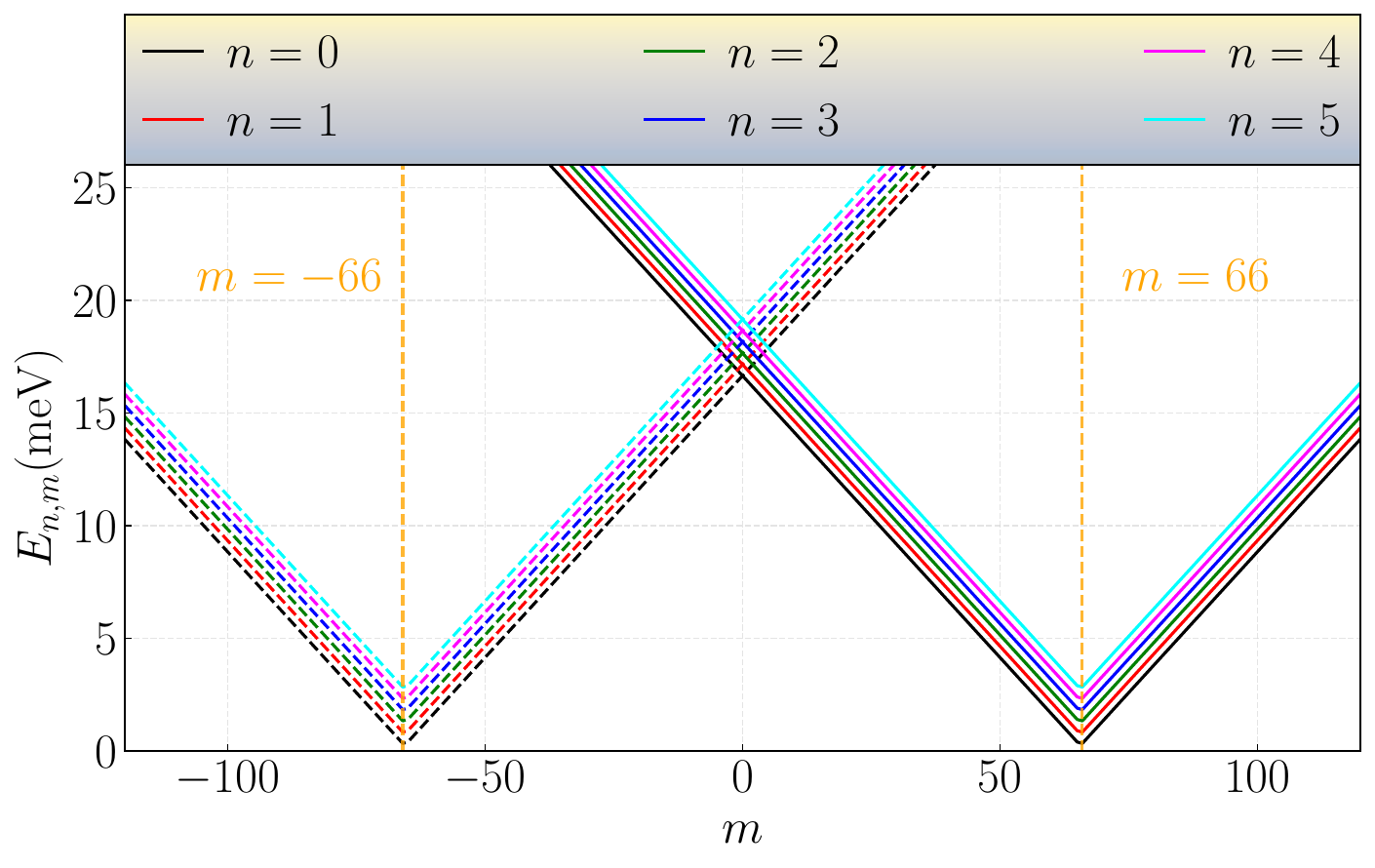}
\caption{\footnotesize 
Energy spectrum $E_{n,m}$ as a function of the angular momentum quantum number $m$, for fixed charge density $\lambda = 0.025$~nC/m and quantum numbers $n = 0,1,\dots,5$.
Each color corresponds to a different value of $n$ ($n=0$ in black, $n=1$ in blue, $n=2$ in green, $n=3$ in red, $n=4$ in orange, and $n=5$ in brown).
The solid curves represent the case $s=+1$ (spin-up), and the dashed curves represent the case $s=-1$ (spin-down).
The mirror-like behavior of the energy levels around $m=0$ is a direct consequence of the AC coupling parameter $\xi$, which changes sign with the spin orientation.}
\label{fig:energy_levels}
\end{figure}
For neutral quantum systems with magnetic dipoles, the polarization profile must account for both the quantum confinement and field-induced alignment:
\begin{equation}
\mathbf{P}_\mu(\mathbf{r}) = \langle \Psi | \hat{\boldsymbol{\mu}} \delta(\mathbf{r} - \mathbf{r}_i) | \Psi \rangle,
\end{equation}
with key validation criteria: (i) System size $L \gg$ de Broglie wavelength $\lambda_d$, (ii) Coherence length $\xi_{top} \gg L$, and (iii)
Temperature $k_B T \ll \hbar\omega_0$.

\begin{table}[h]
\centering
\caption{Comparison of possible realizations}
\begin{tabular}{lll}
\hline
Feature & Generic Model & GaAs Implementation \\
\hline
Particles & Neutral dipoles & Excitons/Dressed electrons \\
Confining potential & Harmonic $V(r)$ & Electrostatic gates \\
Field source & Charged wire & Surface electrodes \\
\hline
\end{tabular}
\label{tab:comparison}
\end{table}

\section{Persistent Current and Hall Conductivity \label{persistent_current}}

To ensure dimensional consistency and extract meaningful physical quantities from the discrete spectrum, we redefine the persistent current associated with a given energy level. From the quantized spectrum
\begin{equation}
E_{n,m} = \hbar \omega_0 \left( 2n + 1 + |m - \xi| \right),\label{energy}
\end{equation}
the current is obtained from the derivative of the energy with respect to the angular momentum quantum number $m$, treated here as a continuous parameter \cite{PRB.1990.42.8351}:
\begin{equation}
I_{n,m} = -\frac{q_{\text{eff}}}{\hbar} \frac{\partial E_{n,m}}{\partial m} = - q_{\text{eff}}\, \omega_0 \cdot \text{sgn}(m - \xi)
\end{equation}
which yields a constant magnitude and sign-reversal around $m = \xi$, consistent with the structure of the spectrum. This expression has the correct physical units of electric current $[\text{A}]$.

To estimate the total current, we assume that $(n + 1)$ energy levels contribute equally, as in standard Landau-level filling arguments. The net current becomes
\begin{equation}
I = - (n + 1) q_{\text{eff}}\, \omega_0 \cdot \text{sgn}(m - \xi).
\end{equation}

In order to define a Hall-like conductivity from this current, we introduce an effective Hall voltage. Since our model does not include an explicit transverse electric field, we adopt a phenomenological approach based on the energy level structure.

To estimate the Hall voltage from the energy spectrum, we consider the energy difference between two adjacent angular momentum states that are symmetrically positioned with respect to the effective coupling point $m = \xi$. Let us define this displacement as $\delta$, such that we compare the energies at $m = \xi + \delta$ and $m = \xi - \delta$, with $\delta$ being a positive integer. This yields $\Delta E_{n,m} = E_{n,\xi + \delta} - E_{n,\xi - \delta}= 2 \hbar \omega_0$. Here, although $\delta$ appears in the intermediate steps, the final result is independent of its value due to the symmetric structure of the spectrum around $\xi$. The energy difference $\Delta E_{n,m} = 2\hbar\omega_0 $ provides a natural scale to define the Hall voltage \cite{PRB.1990.42.8351}. By associating this energy gap with the effective voltage drop in the system, we define the Hall voltage as
\begin{equation}
V^{\text{eff}}_{\text{Hall}} = \frac{\Delta E}{\mu_B} = \frac{2\hbar\omega_0}{\mu_B}
\end{equation}
which can also be interpreted as the effective potential drop between adjacent angular channels.

Substituting this into the conductivity formula $\sigma = I / V_{\text{Hall}}$, we obtain
\begin{equation}
\sigma_{\text{Hall}} = -\frac{\mu_B^2}{h} (n + 1) \cdot \text{sgn}(m - \xi)
\end{equation}
showing that the Hall conductivity is quantized in units of $\mu_B^2/h$ and controlled by the relative sign between $m$ and the AC parameter $\xi$. This result reveals a spin-dependent quantized response in a purely electric-field-driven system of neutral particles, analogous to the integer quantum Hall effect.

For cold atoms, $\mu_B$ is the natural magnetic moment unit, and $q_{\text{eff}}$ effectively characterizes the spin-flip transitions, typically proportional to the bare charge $e$ or its effective counterpart in the system. When comparing to $e^2/h$ units, for cold atoms we can consider $q_{\text{eff}} \sim e$ for practical purposes. In GaAs quantum dots with excitons, $\mu_B$ should be replaced by the effective exciton magnetic moment $\mu_{\text{exc}} = \hbar e/m^* c$, where $m^*$ is the reduced mass.

\begin{figure}[tbh]
\centering
\includegraphics[width=1.0\linewidth]{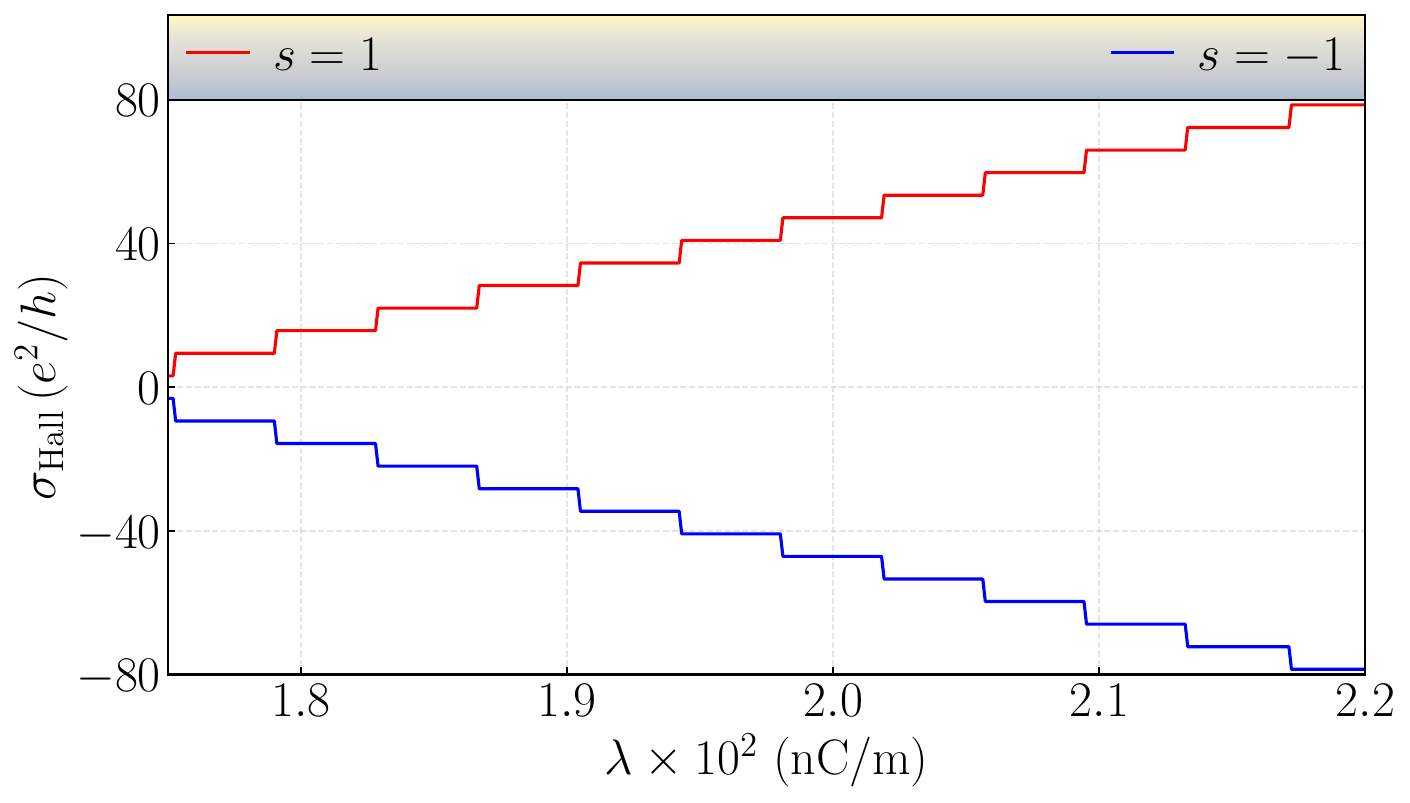}
\caption{\footnotesize Quantum Hall-like conductivity in units of $\mu_B^2/h$ (left axis) and $e^2/h$ (right axis) as a function of the scaled linear charge density $\lambda \times 10^2$ (in nC/m), computed for the spin projections $s = +1$ (green) and $s = -1$ (red). The spin inversion reflects a mirrored structure of Hall plateaus with respect to the axis $\sigma_{\text{Hall}} = 0$.}
\label{fig:hall_conductivity}
\end{figure}

The presence of the sign function also reflects the mirror symmetry of the energy spectrum, as seen in Fig.~\ref{fig:energy_levels}, where the spectrum for $s = -1$ is the mirror image of the case $s = +1$ around $m = \xi$. Thus, the Hall conductivity changes sign with the spin, and the quantized plateaus shift accordingly (Figs.~\ref{fig:energy_levels} and \ref{fig:hall_conductivity}). This shows explicitly how the sign of the Hall conductivity is governed by the relative position of $m$ and $\xi$, and ultimately by the spin orientation through $\mu$.

The linear charge density $\lambda$, which determines the strength of the radial electric field in our model, plays a crucial role in modulating the effective gauge field through the AC coupling $\xi$. This field configuration can be generated in a realistic setting using a charged wire or a cylindrical electrode at the system's center. One can tune $\lambda$ by varying the applied voltage, as is done in nanoscale capacitor structures \cite{PRB.2000.62.7045}.

To estimate the physical values of $\lambda$, consider that a field of $E = 10^7\,\text{V/m}$ at $r = 100\,\text{nm}$ implies $\lambda = E r /2= 5 \times 10^{-4}\,\text{C/m}$. This is within reach of laboratory setups employing scanning probes or nanowires \cite{PRA.2009.79.063613}.

Alternatively, cold atom experiments can engineer synthetic electric fields and spin-orbit interactions through laser arrangements \cite{NJP.2003.5.56,RMP.2011.83.1523,PRL.2001.87.190402}, thereby enabling the simulation of AC-type couplings without the need for physical charges.

A key feature of the model is the role of the spin degree of freedom in determining the sign of the Hall conductivity. Since the AC coupling parameter is defined as $\xi$, and the magnetic moment $\mu$ is proportional to the spin projection $s = \pm 1$, changing the spin state from $s = +1$ to $s = -1$ effectively reverses the sign of $\xi$. This change leads to a mirrored energy spectrum in the angular momentum index $m$, as the energy levels depend on $|m -\xi|$. Consequently, the persistent current and the resulting Hall conductivity are inverted with respect to the spin state. This behavior is clearly observed in Figs.~\ref{fig:energy_levels} and \ref{fig:hall_conductivity}, where the conductivity profile for $s = -1$ appears as the mirror image of the case $s = +1$, with respect to the axis $\sigma_{\text{Hall}} = 0$. Such symmetry provides a direct mechanism for spin-controlled transport in neutral systems, and it could serve as an experimental signature for detecting spin polarization via Hall-like measurements. 
\begin{table}[h]
\centering
\label{tab:param_comp}
\begin{tabular}{lcccl}
\toprule
System & $\lambda\;(10^{-11}\,\mathrm{C/m})$ & $\hbar \omega_{0}\;(\text{meV})$ & Platform & Ref. \\
\midrule
This work & 2.5 $\pm$ 0.2 & 0.25 $\pm$ 0.03 & GaAs QD & -- \\
Photonic systems & 0.5--2.0 & 0.05--0.2 & SiN & \cite{Nature.2019.565.622} \\
Polar molecules & 3.0--8.0 & 0.4--1.2 & NaK & \cite{PRL.2014.113.025301} \\
Electrostatic QDs & 1.0--3.5 & 0.1--0.3 & GaAs & \cite{PRB.2000.62.7045}\\
Cold atoms & 5.0--10.0 & 0.3--1.0 & ${}^{87}$Rb & \cite{PRA.2009.79.063613} \\
\bottomrule
\end{tabular}
\caption{\footnotesize Comparison of experimentally accessible parameters for various quantum systems, highlighting the feasibility of our proposed model. The table lists typical ranges for the linear charge density ($\lambda$), which defines the strength of the electric field and thus the Aharonov-Casher coupling, and the harmonic confinement energy ($\hbar\omega_0$). Our theoretical values for a GaAs Quantum Dot (QD) system ($\lambda = 2.5 \pm 0.2 \times 10^{-11}\,\mathrm{C/m}$ and $\hbar\omega_0 = 0.25 \pm 0.03\,\text{meV}$) fall well within the established experimental capabilities of existing electrostatic GaAs QDs~\cite{PRB.2000.62.7045} and are also consistent with ranges achieved in cold atomic systems~\cite{PRA.2009.79.063613}. This comparison demonstrates that the conditions required to observe the predicted quantum Hall-like effects are within the reach of current laboratory technologies across various physical platforms.}
\label{tab:parameters}
\end{table}

As shown in Table~\ref{tab:parameters}, our parameters $\lambda = 2.5\times10^{-11}$ C/m and $\hbar\omega_0 = 0.25$ meV fall within experimentally accessible ranges for both semiconductor QDs~\cite{PRB.2000.62.7045} and cold atoms~\cite{PRA.2009.79.063613}, ensuring the physical realizability of our model.
\begin{figure}[ht]
\centering
\includegraphics[width=1\linewidth]{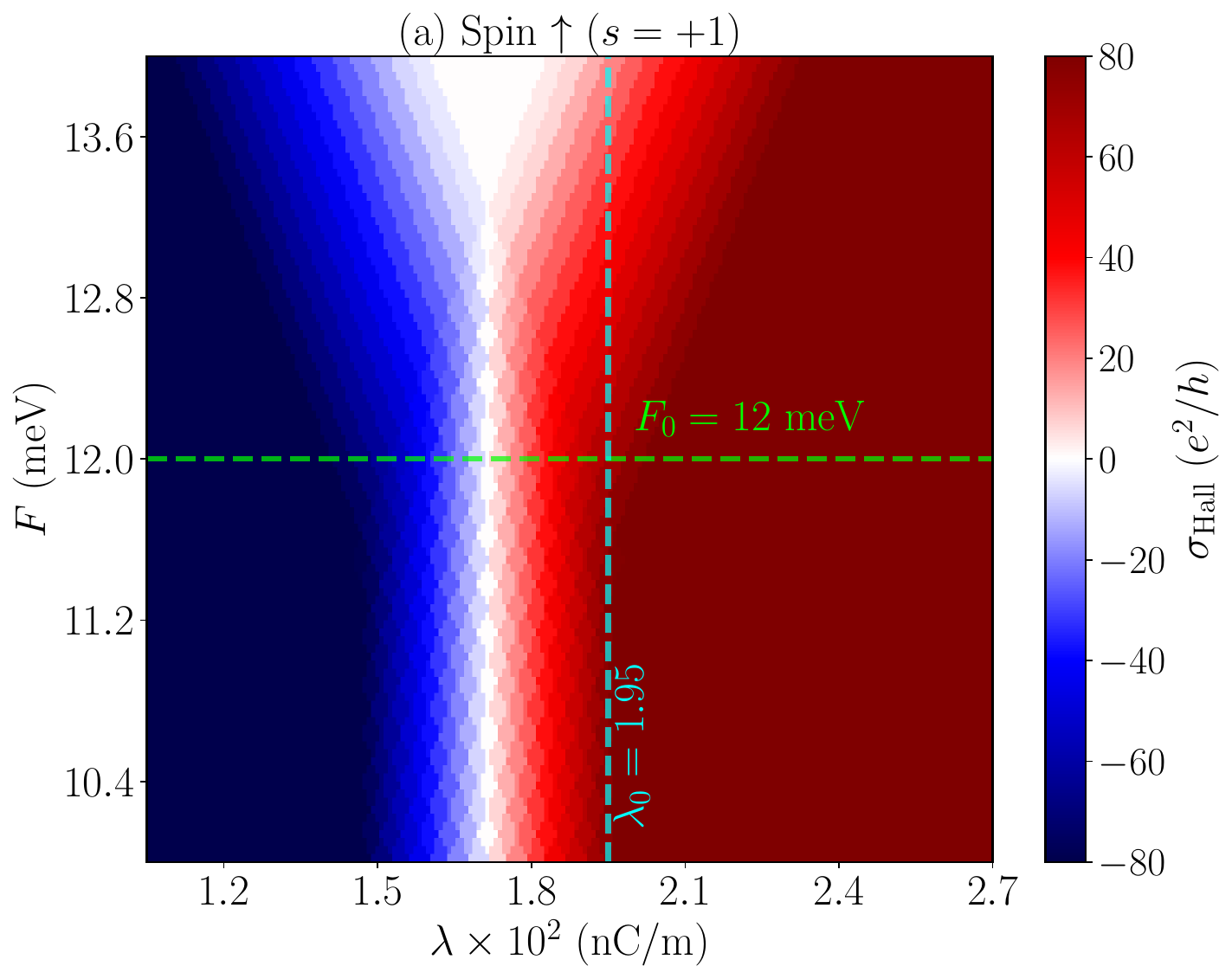} 
\includegraphics[width=1\linewidth]{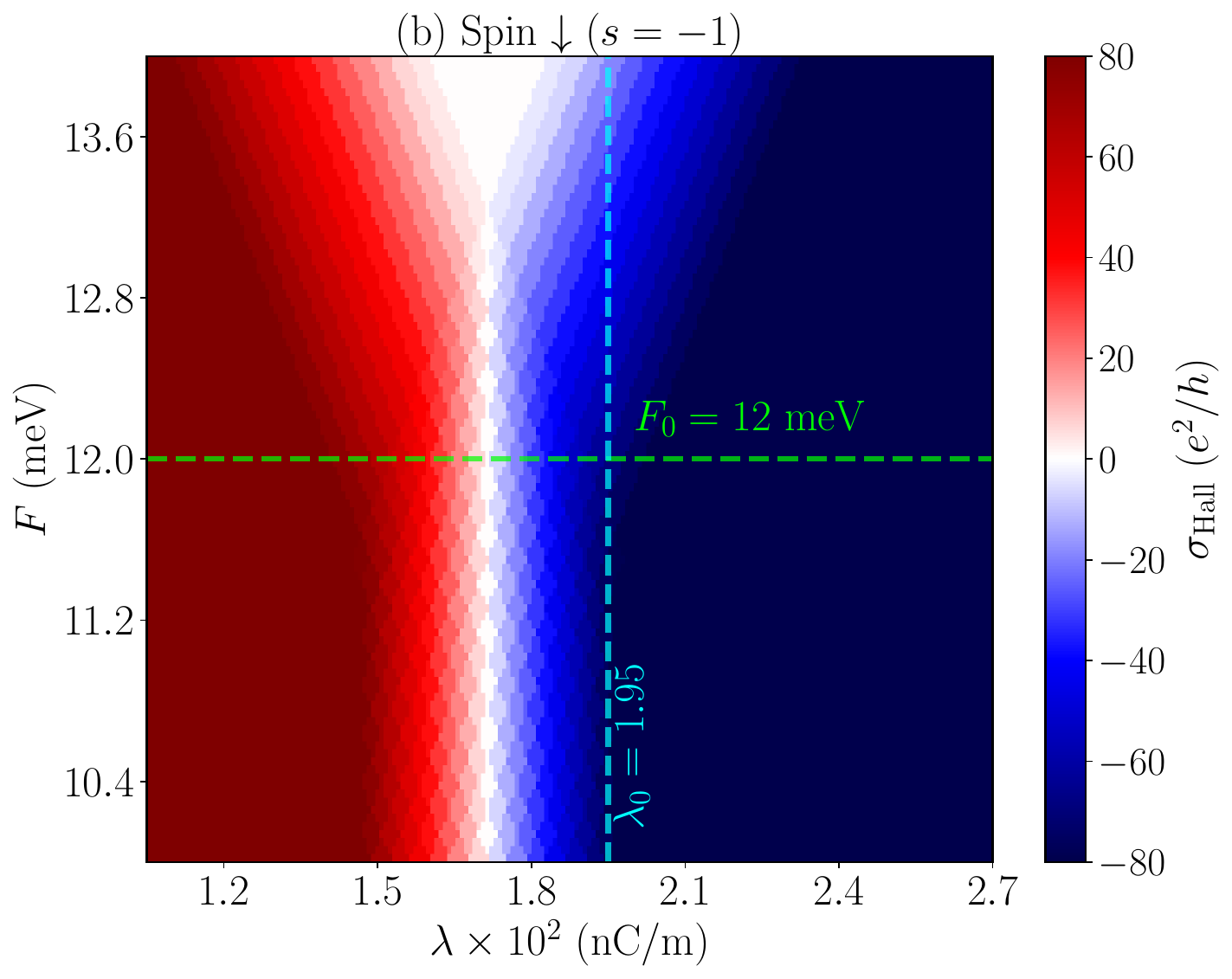} 
\caption{\footnotesize Quantum Hall-like conductivity maps in units of $e^2/h$ for neutral particles with magnetic dipole moments as functions of the scaled linear charge density $\lambda \times 10^2$ (in nC/m) and effective Fermi energy $F$ (in meV). (a) Corresponds to spin-up ($s=+1$), and (b) to spin-down ($s=-1$). The color scale indicates the magnitude and sign of $\sigma_{\mathrm{Hall}}$. The distinct plateaus demonstrate the quantization of conductivity. Note the clear antisymmetric behavior between spin-up and spin-down states, where the sign of the Hall conductivity is inverted upon spin reversal. The dashed lines highlight specific reference values for $F$ ($F_0=12$ meV) and $\lambda$ ($\lambda_0=1.95$) for illustrative purposes, showing cuts where plateaus are prominent. The expanded range of $\lambda$ compared to previous results reveals additional plateau structures in the lower $\lambda$ regime.}
\label{fig:hall_spin_maps_combined}
\end{figure}
%%%%%%%%%%%%%%%%%%%%%%%%%%%%%%%%%%%%%%%%
\begin{figure}[ht]
\centering
\includegraphics[width=1.0\linewidth]{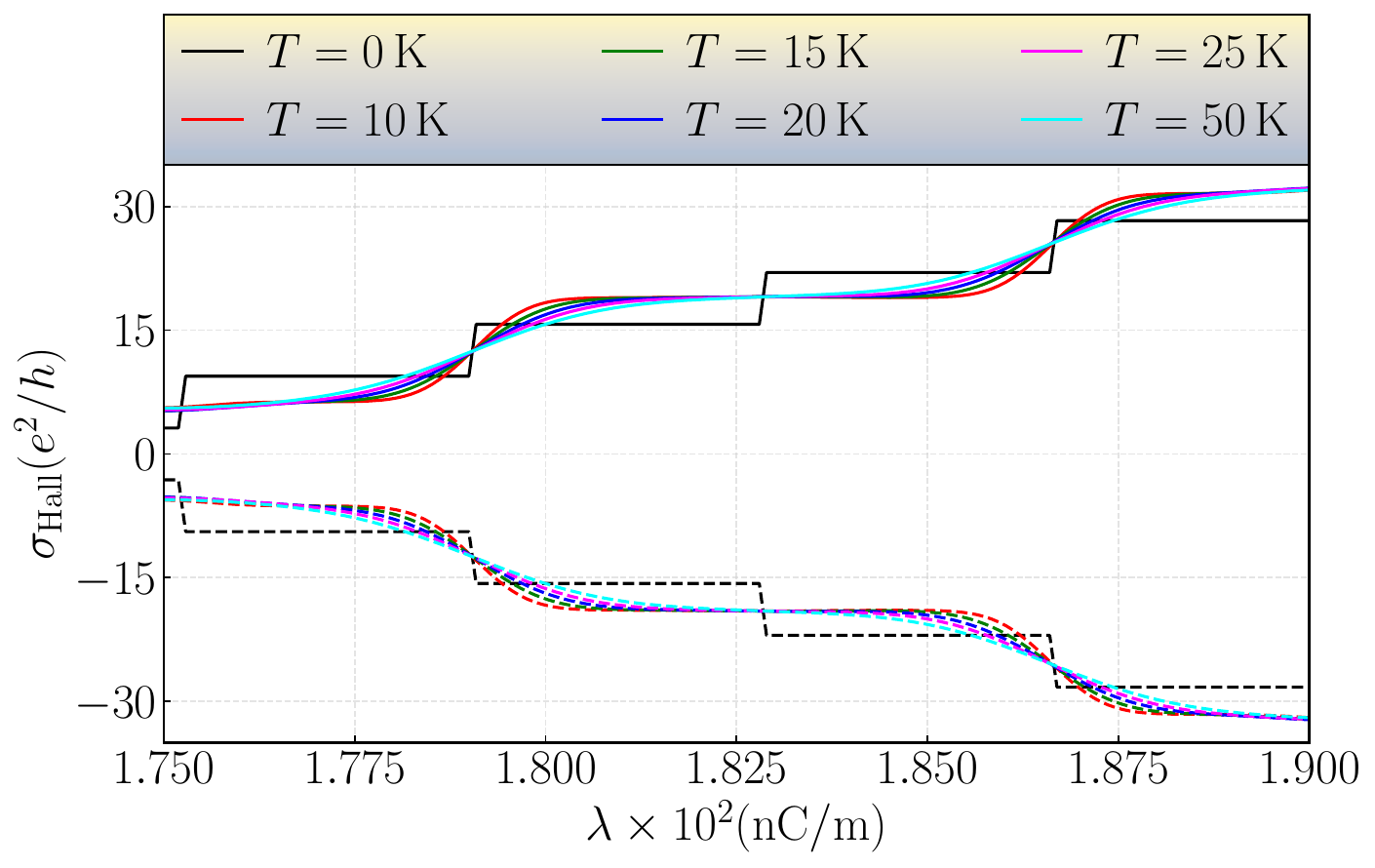}
\caption{\footnotesize Hall conductivity $\sigma_{\mathrm{Hall}}$ (in units of $e^2/h$) as a function of the rescaled linear charge density $\lambda \times 10^2$ (in nC/m), for spin states $s = +1$ (solid lines) and $s = -1$ (dashed lines), at temperatures $T = 0$ (black), $10\,\mathrm{K}$ (orange), $20\,\mathrm{K}$ (blue), $25\,\mathrm{K}$ (red), and $50\,\mathrm{K}$ (purple). The plateaus observed at $T=0$ become increasingly smooth with temperature due to thermal occupation of nearby energy levels. The curves for opposite spins are mirrored with respect to the axis $\sigma_{\text{Hall}} = 0$, revealing a spin-dependent Hall response.}
\label{fig:hall_temp_spin}
\end{figure}
%%%%%%%%%%%%%%%%%%%%%%%%%%%%%
Figure~\ref{fig:hall_spin_maps_combined} presents two-dimensional density maps illustrating the quantum Hall-like conductivity $\sigma_{\mathrm{Hall}}$ for neutral particles with magnetic dipole moments. Panel (a) shows the results for spin-up ($s=+1$) particles, while panel (b) shows those for spin-down ($s=-1$) particles. Both maps span a parameter space defined by the effective Fermi energy, $F$, on the vertical axis (ranging from $10$ meV to $14$ meV), and the scaled linear charge density, $\lambda \times 10^2$, on the horizontal axis (ranging from $1.05$ nC/m to $2.7$ nC/m, with $200$ points). The color intensity and hue in each map directly represent the magnitude and sign of the Hall conductivity, quantized in units of $e^2/h$.

A key feature observed in both panels is the emergence of well-defined, extended regions of uniform color, indicating discrete plateaus in the Hall conductivity. These plateaus are a hallmark signature of quantized transport, analogous to the integer quantum Hall effect. The expansion of the $\lambda$ range to lower values in these new results reveals additional plateau structures, particularly evident in the region of $\lambda \times 10^2 < 1.75$ nC/m, which were not as prominent or visible in previous investigations. The existence of such broad plateaus across variations in both $F$ and $\lambda$ highlights the remarkable robustness of the predicted phenomenon, which is crucial for potential experimental realization.

Crucially, a striking antisymmetric behavior is evident when comparing (a) to panel (b) in Figure~\ref{fig:hall_spin_maps_combined}. For any given region in the parameter space, a positive Hall conductivity for spin-up particles corresponds to a negative conductivity of similar magnitude for spin-down particles, and vice versa. This direct sign reversal of $\sigma_{\mathrm{Hall}}$ upon spin inversion is a profound manifestation of the spin-dependent nature of the Aharonov-Casher coupling parameter $\xi$. It provides a clear mechanism for spin-controlled transport in a purely electric-field-driven neutral system, suggesting avenues for spintronic applications that do not require charge carriers or external magnetic fields.

The dashed lines in both figures, at $F_0 = 12$ meV and $\lambda_0 = 1.95$ (corresponding to $\lambda = 0.0195$ nC/m), serve as visual guides. They delineate specific cuts through the parameter space where the quantized plateaus are particularly pronounced, illustrating how distinct Hall conductivity values can be accessed by tuning these experimentally controllable parameters.

\begin{figure}[h]
\centering
\includegraphics[width=1\linewidth]{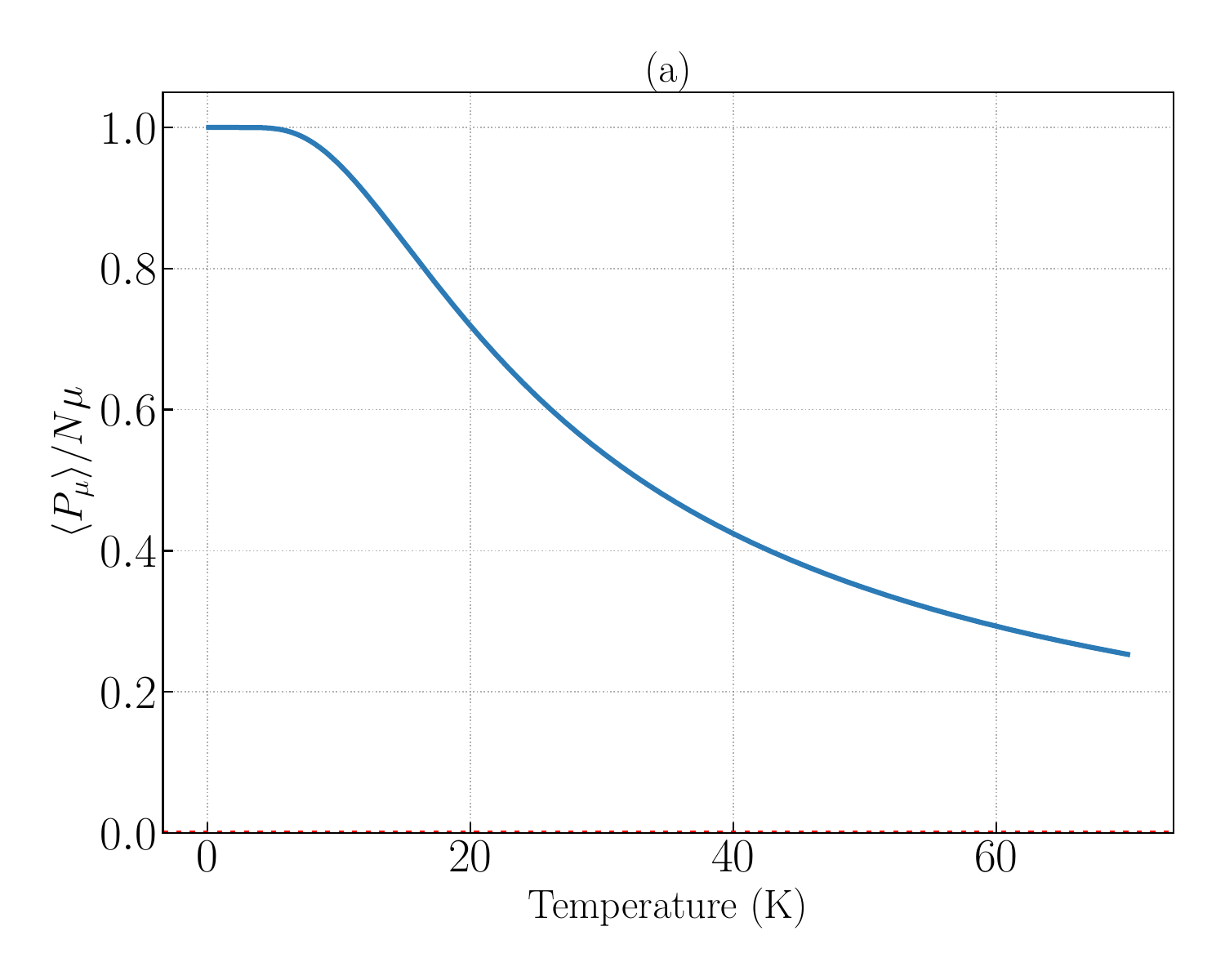}
\includegraphics[width=1\linewidth]{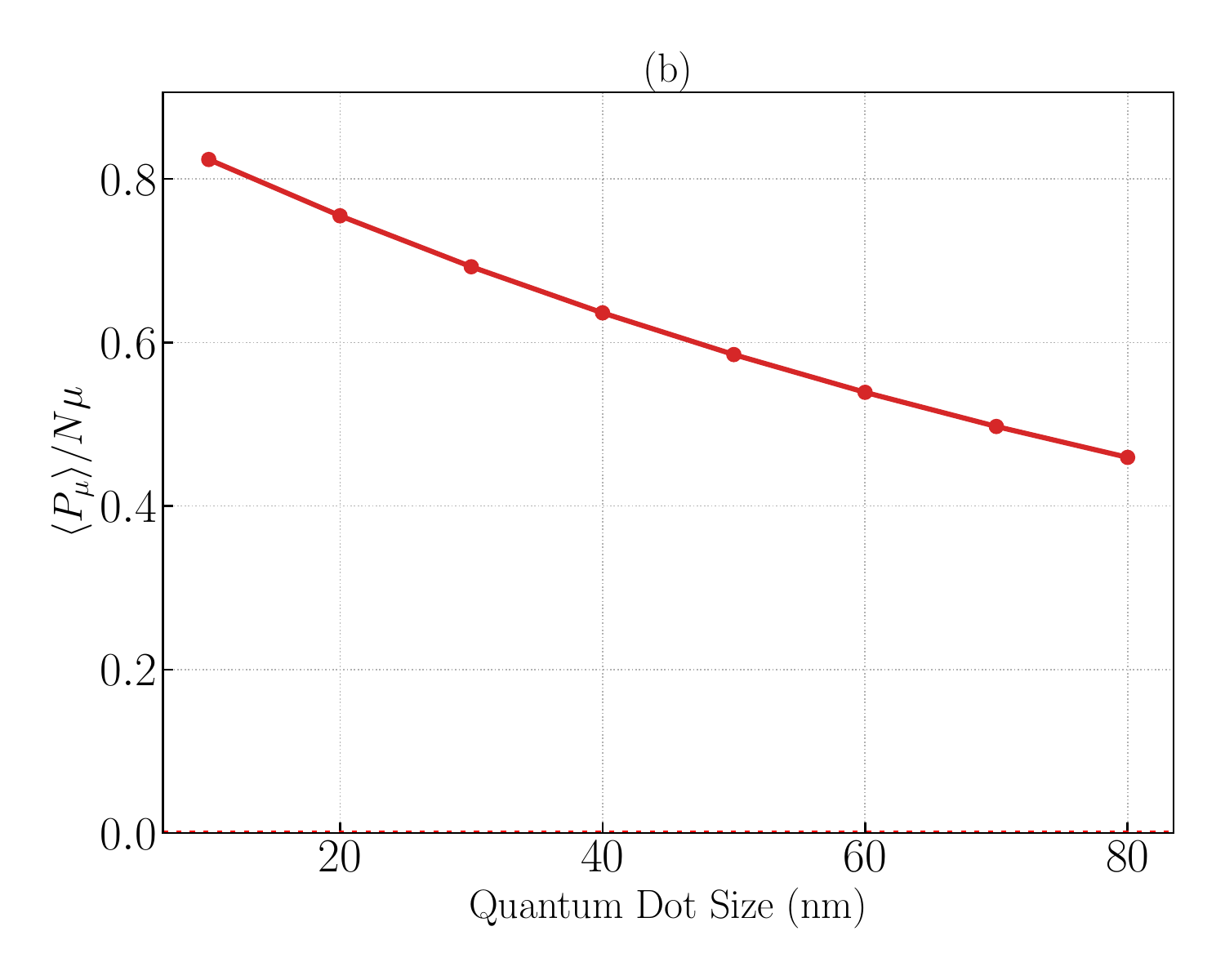}
\caption{\footnotesize Spin Polarization of Neutral Dipoles. (a) Normalized magnetic polarization $\langle P_\mu \rangle / N\mu$ as a function of temperature, calculated using thermodynamic principles. This plot illustrates how the interplay between the Aharonov-Casher-induced spin splitting and thermal fluctuations results in high polarization at low temperatures, which decays as the temperature increases. (b) Normalized magnetic polarization as a function of the quantum dot size (illustrative data). This panel suggests how the spatial confinement or other size-dependent effects could influence the overall spin response, highlighting the potential for device geometry optimization. Together, these plots underscore the critical factors of temperature and geometry in controlling the spin polarization crucial for the quantum Hall-like effect.}
\label{fig:magnetic_polarization_combined}
\end{figure}
%%%%%%%%%%%%%%%%%%%%%%%%%%%%%
Figure~\ref{fig:hall_temp_spin} reveals how temperature affects the Hall-like response of neutral particles under the AC interaction. At absolute zero, the system exhibits clear quantized plateaus in the Hall conductivity, a direct consequence of the discrete energy spectrum resulting from the interplay between the confining potential and the spin-electric field coupling. These plateaus indicate that the current is carried by a well-defined number of quantum states, a hallmark of quantized transport.

As the temperature increases, the sharpness of these steps gradually fades. This smoothing occurs because thermal energy allows particles to populate neighboring levels beyond the sharp Fermi boundary present at zero temperature. Consequently, the distinct transitions between plateaus become less pronounced, and the conductivity acquires a continuous character.

Additionally, the figure illustrates a striking symmetry between spin-up ($s = +1$) and spin-down ($s = -1$) particles. The curves for opposite spins are mirrored with respect to the horizontal axis ($\sigma_{\text{Hall}} = 0$), reinforcing the idea that reversing the spin flips the direction of the induced current, a direct manifestation of the spin-dependence embedded in the AC coupling. This antisymmetric behavior suggests that spin can be used as a control parameter to manipulate transport properties in neutral systems, offering a promising route for spintronic applications in platforms without charge carriers.

\paragraph*{Spin Polarization and Temperature Dependence.} Beyond the quantum Hall-like conductivity, understanding the behavior of the neutral magnetic dipoles in our system requires examining their spin polarization, especially at finite temperatures. The interplay between the energy splitting induced by the AC effect and thermal fluctuations dictates the population distribution between spin states, directly impacting the overall magnetic polarization.

For a system of neutral magnetic dipoles in an effective electric field, the two spin states (spin-up, $N_+$, and spin-down, $N_-$) have distinct interaction energies, leading to an energy splitting $\Delta E$. At a given temperature $T$, their populations are governed by Boltzmann statistics~\cite{KittelThermalPhysics, ReifStatisticalPhysics}. The normalized magnetic polarization $P$ is then defined as the difference in populations divided by the total population, which can be expressed as:
\begin{equation}
P(T) = \frac{N_+ - N_-}{N_+ + N_-} = \tanh\left(\frac{\Delta E}{2k_B T}\right),
\label{eq:polarization_tanh}
\end{equation}
where $k_B$ is the Boltzmann constant. Here, $\Delta E$ represents the effective energy splitting between the spin states induced by the AC interaction.

Figure~\ref{fig:magnetic_polarization_combined}(a) illustrates the normalized magnetic polarization as a function of temperature. At very low temperatures, the thermal energy ($k_B T$) is much smaller than the spin energy splitting ($\Delta E$). This allows the majority of the dipoles to align with the effective field, resulting in a high magnetic polarization (approaching 1). As the temperature increases, however, thermal fluctuations become more significant, progressively overcoming the energy splitting. This causes the spins to randomize, leading to a monotonic decrease in the overall polarization, which eventually approaches zero at high temperatures. This decay of spin polarization fundamentally explains the smoothing and eventual disappearance of the quantized Hall plateaus observed in Figure~\ref{fig:hall_temp_spin}, as thermal occupation of multiple angular momentum channels averages out the quantized current contributions. The persistence of the quantum Hall-like effect up to 25 K, as shown in our previous results, implies that the effective energy splitting $\Delta E$ for the dipoles remains sufficiently large to maintain a non-negligible spin polarization within this temperature range, a critical factor for experimental viability.

Furthermore, the overall spin polarization can also be influenced by the physical dimensions of the quantum dot. Spatial confinement can affect the effective interaction strength or modify the allowed energy levels, leading to size-dependent polarization. In quantum dots, the strength of quantum confinement and particle interactions (including spin-related splittings) is known to depend sensitively on their size and shape \cite{PB.2023.658.414846}. For instance, studies have shown that optical polarization anisotropy and exciton exchange splittings exhibit a clear dependence on quantum dot dimensions, a consequence of how confinement alters wavefunctions and interaction energies[cite: 1, 654]. Figure~\ref{fig:magnetic_polarization_combined}(b) illustrates how the normalized magnetic polarization could potentially vary with the characteristic size of the quantum dot. This dependence, if accurately modeled, suggests that optimizing the device geometry is essential for maximizing the spin response and, consequently, the robustness of the quantum Hall-like effect. Such insights would be particularly relevant for experimental designs in semiconductor quantum dots, where the dot size can be precisely controlled via electrostatic gates.

\section{Conclusion \label{conclusion}}

We have developed a comprehensive theoretical framework for quantum Hall-like effects in neutral particles with magnetic dipole moments subjected to radial electric fields. Our analysis demonstrates that the Aharonov-Casher interaction, when combined with harmonic confinement, produces Landau-level analogs with spin-dependent energy spectra as shown in Eq. (\ref{energy}). The characteristic quantization of Hall conductivity in units of $e^2/h$ emerges naturally from this configuration, despite the absence of charge or external magnetic fields.

The self-adjoint extension treatment of the singularity at the origin, while mathematically necessary for the Hamiltonian's well-posedness, does not significantly impact the physical results for systems with multiple angular momentum channels ($|m| \gg 1$). This is because the Hall conductivity calculation inherently involves summation over many $m$ states, effectively averaging out boundary effects from the low-angular-momentum sector.

Crucially, the detailed two-dimensional maps of Hall conductivity (Figure~\ref{fig:hall_spin_maps_combined}) vividly illustrate the remarkable robustness of this predicted phenomenon. These maps reveal broad and well-defined plateaus of quantized conductivity across extended regions of both the effective Fermi energy and the linear charge density. The expansion of our analysis to lower $\lambda$ values, in particular, uncovers additional plateau structures, further solidifying the stability of the Hall-like response against parameter variations. Moreover, the striking antisymmetric behavior observed upon spin reversal, where the sign of $\sigma_{\mathrm{Hall}}$ flips directly from spin-up to spin-down states, provides a clear and unambiguous experimental signature. This direct spin-control over the transport direction, achieved solely through electric field engineering, opens novel avenues for spintronic applications in neutral systems, circumventing the need for traditional magnetic fields or charge carriers.

This work establishes a fundamental mechanism for topological phenomena in charge-neutral systems, with potential extensions to molecular dipoles or exotic quasiparticles. The spin-dependent quantization revealed in Fig. \ref{fig:hall_conductivity} suggests new possibilities for controlling neutral particle transport through electric field engineering alone. The model proposed in this work demonstrates robust physical and mathematical construction, combining well-established frameworks, such as the Aharonov-Casher effect and self-adjoint extension methods, to address an innovative problem: quantum Hall-like transport in neutral systems under radial electric fields. The consistency of the model is evidenced by the careful resolution of the singularity at the origin, which does not compromise the relevant physical results, particularly in systems with multiple angular momentum channels. Furthermore, the emergent quantization of the Hall conductivity, which is controlled solely by electric fields and spin orientation, represents a significant achievement, as it eliminates the need for external magnetic fields or charged particles. These aspects, along with the experimental feasibility in platforms such as GaAs quantum dots and cold atomic systems, highlight the pioneering nature and practical applicability of this research, positioning it as a valuable contribution to the development of new paradigms in condensed matter physics and topological systems.

Crucially, our predictions operate within parameter ranges accessible to existing experimental platforms. The required confinement energy $\hbar\omega_0 \sim 0.25$ meV matches gate-defined quantum dots, while the electric field strengths correspond to achievable linear charge densities in both semiconductor nanostructures and cold-atom systems. The temperature resilience shown in Fig. \ref{fig:hall_temp_spin} further supports the experimental relevance of these effects.

\section*{Acknowledgments}

This work was supported by CAPES (Finance Code 001), CNPq (Grant 306308/2022-3), and FAPEMA (Grants UNIVERSAL-06395/22 and APP-12256/22).

\bibliographystyle{apsrev4-2}

\end{document}